\documentclass[11pt,a4paper]{article}
\pdfoutput=1

\usepackage[utf8]{inputenc}
\usepackage[english]{babel}
\usepackage{chicago}

\usepackage[round]{natbib}
\usepackage[margin=1in]{geometry}

\usepackage{graphicx, subfig, caption, a4wide, booktabs, fancyvrb, upquote, longtable, multirow, enumitem, ifsym, hyperref}
\usepackage{amsmath, amsfonts, amssymb}
\usepackage{bm}
\usepackage{url}

\usepackage{afterpage}

\usepackage{color}
\definecolor{darkgrey}{rgb}{0.2,0.2,0.2}
\usepackage{fancyvrb}
\DefineVerbatimEnvironment{Code}{Verbatim}
  {fontsize=\normalsize,
   formatcom=\color{darkgrey},
   baselinestretch=1}
   
\makeatletter
\newcommand\code{\bgroup\@makeother\_\@makeother\~\@makeother\$\@codex}
\def\@codex#1{{\normalfont\ttfamily\hyphenchar\font=-1 #1}\egroup}
\makeatother

\usepackage{fancyvrb}
\DefineVerbatimEnvironment{Code}{Verbatim}
  {fontsize=\footnotesize,
   formatcom=\color{darkgrey},
   baselinestretch=0.9}
   
\begin{document}


\title{A Posterior Probability Approach for Gene Regulatory Network
  Inference in Genetic Perturbation Data}
\author{%
  William Chad Young \\ University of Washington \and
  Ka Yee Yeung \\ University of Washington Tacoma \and
  Adrian E. Raftery \\ University of Washington
  \thanks{William Chad Young is a PhD student,
    Department of Statistics, University of Washington, Box 354322,
    Seattle, WA 98195-4322, USA; Email: wmchad@uw.edu. Ka Yee Yeung is
    Associate Professor, Institute of Technology, University of
    Washington - Tacoma, Campus Box 358426 1900 Commerce Street
    Tacoma, WA 98402, USA; Email: kayee@uw.edu. Adrian E. Raftery is
    Professor of Statistics and Sociology, Department of Statistics,
    University of Washington, Box 354322, Seattle, WA 98195-4322, USA;
    Email:  raftery@uw.edu. This research was supported by NIH grants
    U54-HL127624, R01-HD054511 and R01-HD070936. Computational
    resources provided by Microsoft Azure. The authors thank Ling-Hong
    Hung, Mario Medvedovic and Aravind Subramanium for helpful
    discussions.}}  \date{\today}
\maketitle

\begin{abstract}
  Inferring gene regulatory networks is an important problem in
  systems biology. However, these networks can be hard to infer from
  experimental data because of the inherent variability in biological
  data as well as the large number of genes involved. We propose a
  fast, simple method for inferring regulatory relationships between
  genes from knockdown experiments in the NIH LINCS dataset by
  calculating posterior probabilities, incorporating prior
  information. We show that the method is able to find previously
  identified edges from TRANSFAC and JASPAR and discuss the merits and
  limitations of this approach. \\

\end{abstract}

\newpage
\baselineskip=18pt


\section{Introduction}

Gene regulatory networks are very important in understanding the
biological functioning of cells. Identifying the interactions between
genes can aid biologists in their attempts to understand how the cell
functions both in steady state and in reaction to external
stimuli. Unfortunately, due to the complexity of the cells and the
large number of genes involved, discovering the true networks is very
difficult. In most cases the number of genes measured far exceeds the
number of observations, as is typical in microarray or sequencing
experiments. Any method for analyzing such data must take this fact
into account. Often this is done by enforcing a sparsity constraint,
either via an added penalty on non-sparseness or via priors placed on
the model. Even with these constraints, the ability to make valid
inference is limited in these high-dimensional regimes as the number
of genes grows compared to the number of observations
\citep{Wainwright09, Verzelen12}.

Many methods have been developed for inferring relationships between
genes from gene expression data. One approach is to model the network
holistically using Bayesian networks
\citep{Murphy99,Kim03,Kim04,Zou05,Lebre10, Scutari10}. This yields good
interpretability, but often does not scale well and is difficult to
apply at the whole-genome level. Regression-based methods, where the
expression level of a target gene is modeled as a function of the
exression level of that gene's regulators, can be applied to much
larger sets of genes but lack the same overarching model of the entire
network. Inference for these models generally becomes a statistical
variable selection or model selection problem. Common methods for this
include Significance Analysis of Microarrays \citep{Tusher01}, Least
Absolute Shrinkage and Selection Operator (LASSO) \citep{Tibshirani96,
  Gustafsson09, Menendez10} and Bayesian Model Averaging (BMA)
\citep{Raftery97, Hoeting99, Yeung11, Lo12, Young14}.  Another class of
methods looks at mutual information among the measured genes
\citep{Basso05, Margolin06, Faith07, Meyer07}.

When looking for regulatory relationships, it has been found that the
information available from knockout experiments, where a single gene
is fully suppressed, can be highly informative since they give a way
to identify a causal pathway, direct or indirect. In the DREAM4 in
silico network challenge
(http://wiki.c2b2.columbia.edu/dream/index.php?title=D4c2)
\citep{Marbach09, Marbach10}, the winning method used only the data
from the knockout experiments to infer the true networks, ignoring the
time-series data entirely \citep{Pinna10}. In real biological
experiments, full knockout experiments are not possible for many
essential genes, but knockdown experiments, where the target gene is
partially suppressed, are often available. Methods for analyzing
knockdown or knockout data include correlation-based approaches
\citep{Salleh15}, implicit latent variable scoring \citep{Yoo02}, and
Bayesian network scoring \citep{Friedman00, Rogers05, Frohlich07}. In
addition, there has been some work in combining steady-state data with
knockdown data to improve results \citep{Shojaie14, Christley09}.

In this paper, we propose a simple, fast method for inferring gene
regulatory relationships from just knockdown data. Our method uses a
simple linear regression model focusing on single regulator-target
gene pairs based on knockdown data. This method allows the
incorporation of prior knowledge about the relationships and generates
posterior probabilities which can be used to form a ranked edgelist or
as a part of a more expansive analysis.

\section{Data}

Our data come from the National Institute of Health (NIH) Library of
Integrated Network-based Cellular Signatures (LINCS) program
(http://lincsproject.org/) \citep{Duan14}. The aim of this program is
to generate genetic and molecular signatures of cells in response to
various perturbations. One thrust of this program is the large-scale
generation of gene expression profiles using L1000 technology. This
technology has resulted in measurements from over one million
experiments to date on over fifty human cell lines. These cell lines
are populations of cells descended from an original source cell and
having the same genetic makeup, kept alive by growing them in a
culture separate from their original source.

Each of the L1000 experiments measures the expression levels of 1000
landmark genes in the human genome. These genes were selected
specifically to cover as much of the gene expression variation in
cellular expression as possible, since all 20,000+ genes cannot be
measured. These experiments have measured cellular response to more
than 20,000 different chemical perturbagens. In addition, knockdown
and over-expression experiments, where a single gene is targeted to
control its expression level, have been performed on thousands of
individual genes, both within and outside of the 1000 landmark genes.

The L1000 experiments were performed using Luminex Bead technology
\citep{Dunbar06}, in which color-coded microspheres were coded to
attach to specific RNA sequences corresponding to a landmark gene and
fluoresce according to the level of that gene's expression. Sets of
beads for measuring the 1000 landmark genes were added to the solution
for a single experiment along with the perturbing agent. The
experiment was left for a specified period of time and then the gene
expression levels were measured.

Experiments were done in sets on a single plate having individual
wells for 384 experiments. This minimizes some external sources of
error, such as environmental conditions, across these experiments. A
small set of these experiments were used as controls with no
perturbation. This gives a baseline distribution of expression level
for each gene from which to measure deviations in other experiments. A
common approach in this setup is to look at deviations in the
perturbation experiments from the controls on the same plate, again
recognizing that experiments on the same plate are likely to be more
similar than those on different plates. Multiple plates, typically
three or four, are prepared and analyzed together as a batch. These
plates are prepared as technical replicates, with a given perturbation
being prepared and then put into the same well of each plate. This
gives additional power in removing systematic biases. Any given
perturbation also is performed in multiple different batches,
resulting in biological replicates since the sets of experiments were
not prepared together.

\section{Methods}

We want to use the L1000 data to infer gene regulatory networks. This
means that we need a method for inferring causality. One way to do
this would be to use a causal time-series model, but the L1000 data
include a very small number of time points (drug perturbation
experiments include only one to two time points). Instead, we use
knockdown experiments to identify a single gene as a putative causal
agent. Although this limits the amount of information we can gain from
a single experiment, it allows us to use a straightforward model with
a clearly defined regulatory gene.

When looking at the knockdown experiments, it is important to
understand that not all experiments are equally useful. The efficacy
of the perturbation varies between target genes and even between
experiments for the same target. The experimental setup of the LINCS
data is helpful in identifying these differences. We use the control
experiments on a plate to get an estimate of the normal variability of
a gene. This eliminates some of the variability due to effects we
cannot control or even measure, including differences in experimental
conditions such as ambient temperature and the scientist performing
the experiment, since these are captured in plate-level effects.

To take advantage of this aspect of the LINCS data, we calculate
plate-level $z$-values for each gene in a knockdown experiment. To do
this, we first calculate the baseline mean, $\bar{x}_{hp}$, and
standard deviation, $s_{hp}$, for each gene $h$ across all control
experiments on plate $p$. Then the $z$-value for gene $h$ in knockdown
experiment $i$ on plate $p$ is
\begin{equation*}
  \label{eq:2}
  x^*_{hi} = \frac{x_{hi} - \bar{x}_{hp}}{s_{hp}}.
\end{equation*}

Once we have transformed the data in this way, we use a simple linear
regression model to model the change in a target gene $t$ as dependent
on the change in the knockdown gene $h$:
\begin{eqnarray}
  \label{eq:1}
  x^*_{ti} &=& \beta_0 + \beta_1 x^*_{hi} + \epsilon_i,~~ i=1,\dots,n_h \\
  \epsilon_i &\stackrel{\mbox{iid}}{\sim}& N(0, \sigma^2).
\end{eqnarray}
Here, $n_h$ is the number of available knockdown experiments for gene
$h$. This model specifies a linear relationship between the $z$-score
of the knockdown gene $h$ and that of the target gene $t$. This is a
simplification of the true process underlying the relationship between
genes $h$ and $t$, but it can still be effective for discovering
relationships.

We estimate this model with a Bayesian approach using Zellner's
$g$-prior \citep{Zellner86} for the model parameters $(\beta_0,
\beta_1, \sigma^2)$.
\begin{eqnarray*}
  \label{eq:3}
  \beta_1 | \sigma^2, g &\sim& N\left(0, \frac{g \sigma^2}{\sum_i
      x^{*2}_{hi}}\right), \\ 
  \Pr(\beta_0, \sigma^2) &\propto& 1/\sigma^2.
\end{eqnarray*}
The parameter $g$ specifies the expected size of the regression
parameter $\beta_1$ relative to the standard error of the OLS estimate
of $\beta_1$. The choice $g=1$ indicates that the regression parameter
is expected to be nearly indistinguishable from the noise, and thus
gives a lower bound for $g$. Also, $n_h/g$ is the effective number of
data points in the prior, with $g=n_h$ corresponding to a unit
information prior and giving similar results to BIC. We do not want a
prior that has more spread than a unit information prior
\citep{Raftery99}, and thus we choose $g$ in the range $1 \leq g \leq
n_h$. In this case, we used $g = \sqrt{n_h}$. We have found this to be a
good compromise between the extremes. We found that when we estimated
$g$ using an Expectation-Maximization algorithm \citep{Dempster77,
  Young14}, the estimated value was often close to $\sqrt{n_h}$.

The regression model with the $g$-prior allows us to quickly calculate
the posterior probability that gene $h$ regulates gene $t$
\citep{Clyde04}. We first calculate the ratio of the likelihood that
$h$ regulates $t$ given the data $\mathbf{x}$, $\Pr(h \rightarrow
t|\mathbf{x})$, versus the likelihood that there is no regulatory
relationship, $\Pr_0$. Further, we can incorporate a prior probability
of regulation, $\pi_{ht}$, which reflects prior information regarding
a regulatory relationship between genes $h$ and $t$. This gives us
\begin{equation*}
  \label{eq:6}
  T_{ht} \equiv \frac{\Pr(h \rightarrow t|\mathbf{x})}{\Pr_0} =
  \frac{\pi_{ht}}{1-\pi_{ht}} \exp \left[ (n_h-2)\log(1+g)/2 - (n_h-1)
    \log(1+g(1-R^2))/2 \right], 
\end{equation*}
where $R^2$ is the coefficient of determination for the simple linear
regression model (\ref{eq:1}). From this we can get the posterior
probability that $h$ regulates $t$, or posterior edge probability:
\begin{equation*}
  \label{eq:5}
  p_{ht} = \frac{\Pr(h \rightarrow t|\mathbf{x})}{\Pr(h \rightarrow
    t|\mathbf{x}) + \Pr_0} = \frac{T_{ht}}{1+T_{ht}}.
\end{equation*}
We use this posterior edge probability to rank potential edges and
find likely edges for further investigation.

Two advantages of this method are its speed and simplicity. To compute
the z-scores, we first get plate-level means and standard deviations,
which can be done in a single read through the baseline data by
keeping track of sums and sums of squares. From there, standardization
of the knockdown data is quick and we need only to calculate
correlations between the knockdown gene and each other gene to get the
posterior edge probabilities. Additionally, including external knowledge
through the prior edge probability can provide a significant boost in
accuracy and precision \citep{Young14}. Finally, these posterior
probabilities have a straightforward interpretation, namely the
probability that a given regulator-target pair is a true relationship
given the data.

\section{Results}

We computed posterior probabilities for edges on the LINCS data for
cell line A375. Cell line A375 is a human skin melanoma cell line with
over 100,000 experiments in the LINCS data. That includes
approximately 15,000 knockdown experiments on landmark genes. This
gives a good set of data to evaluate our method. We set the prior
probability of any edge being present to 0.0005, reflecting the
average expected number of regulators (parents) for each node
determined by Guelzim et al. \citep{Guelzim02} for yeast and the
assumption that transcription factors will regulate approximately
the same proportion of target genes regardless of the total number of
genes available.

To assess our results, we need a reference standard. In our case, we
looked at the Enrichr website \citep{Chen13}, which has collected
numerous gene-set libraries, including some that list gene regulatory
relationships. We used the TRANSFAC and JASPAR lists of edges; these
list transcription factors as well as putative binding sites on other
genes using a position weight matrix \citep{Wingender00,
  Sandelin04}. This is not a comprehensive gold standard for
assessment since these regulatory relationships are limited to
well-studied transcription factors. However, assessment of gene
networks in the mammalian systems is non-trivial due to incomplete
knowledge.

The TRANSFAC and JASPAR (T\&J) dataset includes 37 transcription
factors that overlap with the LINCS landmark genes. Thus we limit our
assessment to only those genes as potential regulators. For these 37
transcription factors, the T\&J dataset has approximately 4,200
regulation-target pairs among the landmark genes out of about 42,000
possible pairs for which we have posterior probabilities.

To further evaluate our method, we compared our results with results
from applying Significance Analysis of Microarrays (SAM)
\citep{Tusher01} and mutual information methods to the data. SAM is an
adaptable method for identifying significant changes in gene
expression level while estimating the false discovery rate. It is
widely used to evaluate microarray data and is available as an
\texttt{R} package, \texttt{samr}. Mutual information methods, based
in information theory, have also been used extensively to identify
relationships among genes. We used the \texttt{minet} package in
\texttt{R} \citep{Meyer08} to analyze our data with three different
mutual information methods: Context Likelihood of Relatedness (CLR)
\citep{Faith07}, Algorithm for the Reconstruction of Accurate Cellular
Networks (ARACNE) \citep{Margolin06}, and minimum redundancy - maximum
relevance (MRMR) \citep{Ding05, Meyer07}. These three methods offer
differing approaches for identifying relationships between genes.

Each method produces a list of gene pairs along with some measure of
the strength of their relationship. SAM returns p-values for each
relationship, the mutual information methods produce weights
indicating the strength of the relationship, and our method gives
posterior probabilities. We can sort these to produce a ranked list
and evaluate these lists against the reference dataset.

First, we looked at two-by-two tables from each method. For the
posterior probability method, we used probability cutoffs of 0.5 and
0.95 to define found edges. SAM provides a list of relationships found
to be significant. The mutual information methods do not define a
particular cutoff for significance, and so all relationships returned
with non-zero weight were included. These two-by-two tables are
produced in Table \ref{tab:assessment2x2}. To assess whether the lists
and the T\&J dataset are related, we also report approximate
(non-Bayesian) $p$-values by using the probability of getting at least
the number of true positives found using a binomial($n$, $p$)
distribution, where $n$ is the number of pairs in the inferred list
and $p$ is the probability of selecting a true edge from the total
number of possible edges. Those $p$-values are equal to 0.02 for both
thresholds, indicating that the posterior edge probabilities are
related to the T\&J results at conventional levels of
significance. The competing methods are not able to accurately
identify a small number of edges as true, returning many more than the
posterior probability method.

\begin{table}[ht]
  \centering
  \caption{Assessment results showing 2x2 tables for cell line A375
    using knockdown experiments for finding edges via posterior
    probability calculation and compared to approximately 4,200 edges
    from TRANSFAC and JASPAR across 37 transcription factors. When
    looking at edges with posterior probability of 0.5 or greater
    (top left table), 41 of the 292 candidate edges are found in 
    TRANSFAC and JASPAR, and 14 of the 76 candidate edges at a cutoff
    of 0.95 (top center table) are true edges. Binomial approximate
    p-values are 0.02 (left) and 0.02 (right). The competing methods
    return many more edges as true but are not as precise, resulting
    in lower p-values.}  
  \begin{tabular}[ht]{rlcclcclcc}
    & & \multicolumn{2}{c}{\textbf{PP 0.5}} & \hspace{5mm} &
    \multicolumn{2}{c}{\textbf{PP 0.95}} & \hspace{5mm} &
    \multicolumn{2}{c}{\textbf{SAM}} \\
    & & Yes & No & & Yes & No & & Yes & No\\
    \cline{3-4} \cline{6-7} \cline {9-10}
    \multirow{2}{*}{\textbf{T\&J}} & Yes & \multicolumn{1}{|c|}{41} &
    \multicolumn{1}{c|}{4262} & & \multicolumn{1}{|c|}{14} &
    \multicolumn{1}{c|}{4289} & & \multicolumn{1}{|c|}{1193} &
    \multicolumn{1}{c|}{3110} \\ 
    \cline{3-4} \cline{6-7} \cline {9-10}
    & No & \multicolumn{1}{|c|}{251} & \multicolumn{1}{c|}{37566} & &
    \multicolumn{1}{|c|}{62} & \multicolumn{1}{c|}{37755} & &
    \multicolumn{1}{|c|}{11151} & \multicolumn{1}{c|}{26666} \\
    \cline{3-4} \cline{6-7} \cline{9-10}
    & & \multicolumn{2}{l}{p-value: 0.02} & &
    \multicolumn{2}{l}{p-value: 0.02} & & \multicolumn{2}{l}{p-value:
      0.98} \\ ~ \\
    & & \multicolumn{2}{c}{\textbf{CLR}} & \hspace{5mm} &
    \multicolumn{2}{c}{\textbf{ARACNE}} & \hspace{5mm} &
    \multicolumn{2}{c}{\textbf{MRMR}} \\
    & & Yes & No & & Yes & No & & Yes & No\\
    \cline{3-4} \cline{6-7} \cline {9-10}
    \multirow{2}{*}{\textbf{T\&J}} & Yes & \multicolumn{1}{|c|}{1651} &
    \multicolumn{1}{c|}{2652} & & \multicolumn{1}{|c|}{34} &
    \multicolumn{1}{c|}{4269} & & \multicolumn{1}{|c|}{1530} &
    \multicolumn{1}{c|}{2773} \\ 
    \cline{3-4} \cline{6-7} \cline {9-10}
    & No & \multicolumn{1}{|c|}{14671} & \multicolumn{1}{c|}{23146} & &
    \multicolumn{1}{|c|}{910} & \multicolumn{1}{c|}{36907} & &
    \multicolumn{1}{|c|}{13533} & \multicolumn{1}{c|}{24284} \\
    \cline{3-4} \cline{6-7} \cline{9-10}
    & & \multicolumn{2}{l}{p-value: 0.67} & &
    \multicolumn{2}{l}{p-value: 1.00} & & \multicolumn{2}{l}{p-value:
      0.60} \\
  \end{tabular}
  \label{tab:assessment2x2}
\end{table}

We expect errors from the assessment results in the form of both false
postives and false negatives due not only to limitations of our
method, but also due to the nature of the data and the TRANSFAC and
JASPAR reference standard by which we evaluate our edges. This is in
part because the T\&J reference standard is not specific for the given
cell line A375. Additionally, false positives might arise because the
expression levels of target genes change due to indirect effects. The
path from the transcription factor to the target gene may go through
intermediate genes. In fact, since only about 5\% of the human genes
are measured by the LINCS experiments, there are likely to be many
genes in relevant pathways that are not measured. If we had
measurements for all 22,000 genes, using a link removal procedure
could be very useful \citep{Klamt10, Pinna10}. We also expect false
negatives since the T\&J dataset is not a set of confirmed regulatory
relationships. Rather, it is informed from attributes of the
transcription factor as well as the target gene. This means that many
of the true relationships as designated by the T\&J dataset may not in
fact reflect true interaction at the cellular level. In general, we do
not expect a transcription factor to affect 10\% of the possible
targets, which is what the T\&J dataset reports, so it is likely that
the T\&J data is overestimating the number of regulatory
relationships.

Another way of looking at the results is via the precision-recall
curve. Precision and recall are both calculated by truncating our
ranked list of edges and looking only at those proposed
edges. Precision is the proportion of the proposed edges which are
true edges. Recall is the proportion of true edges which are in the
proposed set. The precision-recall curve takes a ranked list of edges
from a procedure and shows how the precision varies as more and more
edges are included from that list. High precision at low recall
indicates that the procedure is good at identifying true edges at the
highest probability. This is important in many cases, particularly
genetic studies, because it gives researchers good information on
where to focus their efforts in subsequent studies.

\begin{figure}[ht]
  \centering
  \includegraphics[width=\textwidth]{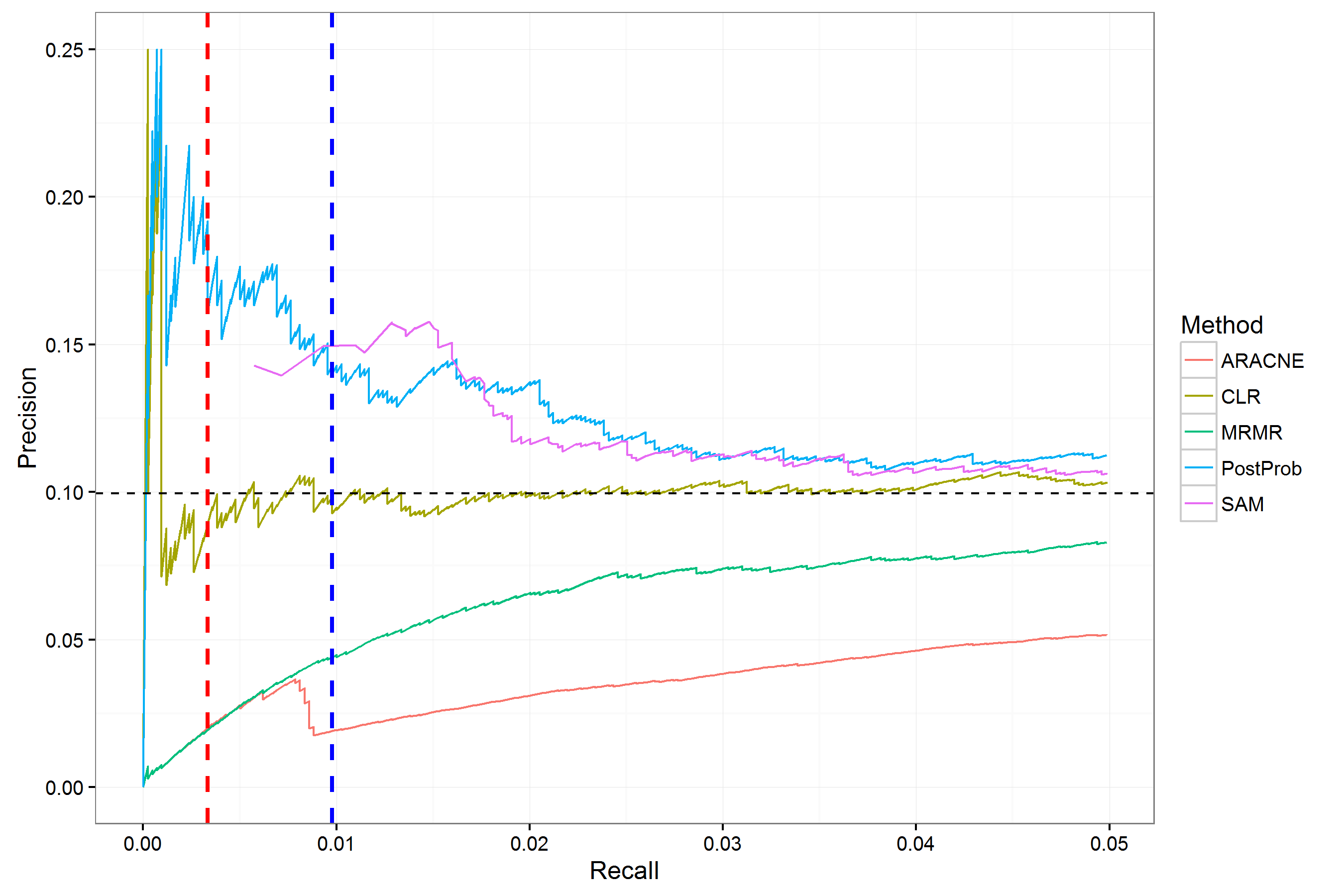}
  \caption{Precision-recall curves for cell line A375 using knockdown
    experiments for finding edges via posterior probability
    calculation evaluated against approximately 4,200 edges from
    TRANSFAC and JASPAR across 37 transcription factors. To the left
    of the red line are those edges with a posterior probability of at
    least 0.95, while the blue line shows the cutoff for edges with a
    posterior probability of 0.5. The horizontal dashed line shows the
    expected precision of 0.1 that would result from randomly ranking
    edges.}
  \label{fig:auprc}
\end{figure}

Figure \ref{fig:auprc} shows the precision-recall curves generated by
the different methods. We do see that the edges most highly ranked by
posterior probability yield better results than expected from random
guessing by a factor of 1.5 to 2. The precision declines as we add
more edges, returning to hover near random guessing. The MRMR and
ARACNE results fare worse than random guessing, and although CLR ranks
a few true edges highly, it returns to random much faster than the
posterior probability edges. The ranked list from SAM performs
comparably to the posterior probability method, but it is unable to
differentiate between the edges at the very top of its list, with 168
edges yielding the same lowest $p$-value. From a scientific point of
view, it is important to have high precision among the edges ranked
most highly, since there are limited resources for designing and
executing experiments investigating particular edges more closely. Of
course it would be preferable to see even better precision, but our
previous discussion has shown why that may not be achievable with this
dataset and standard.

\section{Discussion}

We have demonstrated a straightforward approach to inferring gene
regulatory network edges from knockdown data. This approach is simple
to apply to large datasets and includes the ability to incorporate
prior information when available. This approach is able to find
confirmable regulatory relationships between genes from the L1000
data. We showed that our method performs comparably to or better than
popular approaches for identifying important regulatory relationships
as found in the TRANSFAC \& JASPAR evaluation dataset.

One key benefit of this approach is that it can be applied to
extremely large datasets, requiring only one read through the data to
compile all sufficient statistics for computing the posterior
probabilities. There is no need to retain all the data after reading
it and no iterative methods, such as Expectation-Maximization or
Markov Chain Monte Carlo, are used. Methods which model the entire
network \citep{Lebre10, Scutari10, McGeachie14, Sanchez13} may yield more
comprehensive network results but are also generally restricted to a
smaller set of genes due to computational constraints. Our approach is
complementary to these other approaches in potentially narrowing down
the set of edges for further investigation.

We have applied the method to knockdown data in order to identify
causal regulatory relationships. This method can also be applied to
over-expression data or even steady-state data, although for
steady-state data the resulting edges would lack directionality
\citep{Michailidis13}. This method could also be used to infer
differential expression for a perturbation such as a drug
treatment. This could be done using a 2-class model where the
predictor variable indicates whether the expression measurements come
from a perturbation experiment or a control experiment. An
implementation of our method will be available as an \texttt{r}
package, \texttt{BayesKnockdown}, including functions for both
knockdown and perturbation data.

We considered using an edge reduction technique, such as that used by
Pinna et al. \citep{Pinna10}, but ultimately decided against it. Since
we only used 37 genes as regulators due to our assessment data, the
resulting networks did not tend to have multiple pathways from one
gene to another. In cases where the resulting networks are much more
rich in multi-gene pathways, using an edge reduction method may be
appropriate.

Another possible use of this method is to use the resulting edge
probabilities as an informed prior for another method utilizing a
different type of data. This allows the integration of multiple data
sources and may increase the usefulness of knockdown data that may
expected to only provide a small amount of evidence within a larger
experimental context.

\section*{Acknowledgments}

This research was supported by National Institutes of Health [R01
HD054511 and HD070936 to A.E.R., U54 HL127624 to A.E.R. and K.Y.Y.];
Microsoft Azure for Research Award to K.Y.Y.; and Science Foundation
Ireland ETS Walton visitor award 11/W.1/I207 to A.E.R.

\bibliographystyle{apalike}
\bibliography{references}

\end{document}